%
\input harvmac.tex
%
%
\noblackbox


\def\inbar{\,\vrule height1.5ex width.4pt depth0pt}
\def\IB{\relax{\rm I\kern-.18em B}}
\def\IC{\relax\hbox{$\inbar\kern-.3em{\rm C}$}}
\def\ID{\relax{\rm I\kern-.18em D}}
\def\IE{\relax{\rm I\kern-.18em E}}
\def\IF{\relax{\rm I\kern-.18em F}}
\def\IG{\relax\hbox{$\inbar\kern-.3em{\rm G}$}}
\def\IH{\relax{\rm I\kern-.18em H}}
\def\II{\relax{\rm I\kern-.18em I}}
\def\IK{\relax{\rm I\kern-.18em K}}
\def\IL{\relax{\rm I\kern-.18em L}}
\def\IM{\relax{\rm I\kern-.18em M}}
\def\IN{\relax{\rm I\kern-.18em N}}
\def\IO{\relax\hbox{$\inbar\kern-.3em{\rm O}$}}
\def\IP{\relax{\rm I\kern-.18em P}}
\def\IQ{\relax\hbox{$\inbar\kern-.3em{\rm Q}$}}
\def\IR{\relax{\rm I\kern-.18em R}}
\font\cmss=cmss10 \font\cmsss=cmss10 at 7pt
\def\IZ{\relax\ifmmode\mathchoice
{\hbox{\cmss Z\kern-.4em Z}}{\hbox{\cmss Z\kern-.4em Z}}
{\lower.9pt\hbox{\cmsss Z\kern-.4em Z}}
{\lower1.2pt\hbox{\cmsss Z\kern-.4em Z}}\else{\cmss Z\kern-.4em
Z}\fi}
\def\IGa{\relax\hbox{${\rm I}\kern-.18em\Gamma$}}
\def\IPi{\relax\hbox{${\rm I}\kern-.18em\Pi$}}
\def\ITh{\relax\hbox{$\inbar\kern-.3em\Theta$}}
\def\IOm{\relax\hbox{$\inbar\kern-3.00pt\Omega$}}

\font\zfont = cmss10 
\font\litfont = cmr6

\def\bigone{\hbox{1\kern -.23em {\rm l}}}
\def\ZZ{\hbox{\zfont Z\kern-.4emZ}}

\def\half{{\litfont {1 \over 2}}}

\def\CJ{{\cal J}}

\def\CQ{{\cal Q}}

\def\CN{{\cal N}}

\def\IR{\relax{\rm I\kern-.18em R}}
\def\Dsl{\,\raise.15ex\hbox{/}\mkern-13.5mu D} 
\def\Gsl{\,\raise.15ex\hbox{/}\mkern-13.5mu G} 
\def\Csl{\,\raise.15ex\hbox{/}\mkern-13.5mu C} 
\font\cmss=cmss10 \font\cmsss=cmss10 at 7pt
%
%
\lref\wittfive{E. Witten, ``Five-Brane Effective Action In M-Theory,''
hep-th/9610234.}
\lref\ahaw{O. Aharony and E. Witten, ``Anti-de Sitter Space
and the Center of the Gauge Group,''
hep-th/9807205.}
\lref\afgj{D. Anselmi, D.Z. Freedman, M.T. Grisaru, and A.A. Johansen,
``Nonperturbative formulas for central functions of supersymmetric
gauge theories,''   hep-th/9708042}
\lref\dlm{M.J. Duff, J.T. Liu and R. Minasian,``Eleven dimensional
origin of string/string duality: a one loop test,'' Nucl. Phys. {\bf B452}
(1995) 261, hep-th/9506126.}
\lref\largen{J. Maldacena, ``The Large N Limit of Superconformal Field Theories
and Supergravity,'' hep-th/9711200.}
\lref\wittads{E. Witten, ``Anti de Sitter space and holography,''
hep-th/9802150.}
\lref\hst{P. Howe, G. Sierra, and P. Townsend,
``Supersymmetry in six dimensions,''  Nucl. Phys.
{\bf B221} (1983) 331.}
\lref\ppvn{M. Pernici, K. Pilch and P. van Nieuwenhuizen, ``Gauged maximally
extended supergravity in seven dimensions,''  Phys. Lett. {\bf B143} (1984)
103.}
\lref\krvn{M. Gunaydin, L. J. Romans and N. P. Warner, ``Gauged $N=8$
supergravity in five dimensions,''  Phys. Lett. {\bf B154} (1985) 268\semi H.J.
Kim ,  L.J. Romans, P. van Nieuwenhuizen,
``The mass spectrum of chiral $N=2,D=10$ supergravity
on $S^5$,'' Phys. Rev. {\bf D32} (1985) 389\semi M. Pernici, K. Pilch and P.
van Nieuwenhuizen,
``Gauged $N=8$, $d=5$ supergravity,''  Nucl. Phys. {\bf B259} (1985) 460. }
\lref\fhmm{D. Freed, J.A. Harvey, R. Minasian and G. Moore, ``Gravitational
Anomaly Cancellation for  $M$-Theory Fivebranes,''  hep-th/9803205.}
\lref\bottcatt{R. Bott and A.S. Cattaneo, ``Integral
invariants of 3-manifolds,'' dg-ga/9710001.}
\lref\clash{A. Losev, G. Moore, N. Nekrasov, and
S. Shatashvili, ``Chiral Lagrangians, Anomalies, Supersymmetry,
and holomorphy,'' Nucl. Phys. {\bf B484} (1997)196, hep-th/9606082.}
\lref\klebs{ I.R. Klebanov and A.A. Tseytlin, `` Entropy of Near-Extremal Black
$p$-branes,'' Nucl. Phys. {\bf B475} (1996) 164, hep-th/9604089\semi  S.S.
Gubser and I.R. Klebanov, ``Absorption by Branes
and Schwinger Terms in the World Volume Theory,'' Phys. Lett.
{\bf B413} (1997) 41, hep-th/9708005.}
\lref\hennsken{M. Henningson and K. Skenderis, ``The Holographic Weyl
Anomaly,'' hep-th/9806087.}
\lref\msw{J. Maldacena, A. Strominger and E. Witten, ``Black Hole Entropy
in M Theory,'' J. High Energy Phys. 12 (1997) 2, hep-th/9711053.}
\lref\freund{P.G.O. Freund and M. A. Rubin, ``Dynamics of Dimensional
Reduction,'' Phys. Lett. {\bf 97B} (1980) 233.}
\lref\kkrefs{
M.J. Duff, B.E.W. Nilsson and C.N. Pope, ``Kaluza-Klein Supergravity,''
Phys. Rep. 130 (1986) 1.}
\lref\gkp{S. Gubser, I. Klebanov and A. Polyakov,
``Gauge Theory Correlators
from Noncritical String Theory,'' Phys. Lett. {\bf B428}
(1998) 105, hep-th/9802109.}
\lref\GrHa{P.~ Griffiths and J.~ Harris, {\it Principles of
Algebraic Geometry},  J.Wiley and Sons, 1978. }
\lref\vwrsq{C. Vafa and E. Witten, ``A One-loop Test of String Duality,''
Nucl. Phys. {\bf B447} (1995) 261, hep-th/9505053.}
\lref\cnss{D.Z. Freedman, S.D. Mathur, A. Matusis and L. Rastelli,,
``Correlation functions in the $CFT(d)/AdS(d+1)$ correpondence,''
hep-th/9804058\semi G. Chalmers, H. Nastase, K. Schalm and R. Siebelink,
``R-Current Correlators in $N=4$ Super Yang-Mills Theory from
Anti-de Sitter Supergravity,'' hep-th/9805105.}
\lref\pvnt{K. Pilch, P. van Nieuwenhuizen, P.K. Townsend,
``Compactification of $D=11$ supergravity on
$S(4)$ (or $11=7+4$, too),''  Nucl. Phys.{\bf B242} (1984) 377.}
\lref\acf{{\it Modern
Kaluza-Klein Theories},
T. Appelquist, A. Chodos, and P.G.O. Freund, eds.
Frontiers in Physics v. 65, Addison-Wesley, 1987.}
\lref\greenco{M.B. Green and P. Vanhove, ``D-instantons, Strings and
M-theory,''  Phys.Lett. {\bf B408} (1997) 122, hep-th/9704145\semi M.B. Green,
M. Gutperle and P. Vanhove, ``One loop in eleven dimensions,''  Phys.Lett. {\bf
B409} (1997) 177, hep-th/9706175.}
\lref\afmn{ I. Antoniadis, S. Ferrara, R.~Minasian and K.S. Narain, ``$R^4$
Couplings in M and Type II Theories on Calabi-Yau Spaces,'' Nucl. Phys. {\bf B
507} (1997) 571; hep-th/9707013.}
%
%
%

\Title{\vbox{\baselineskip12pt
\hbox{EFI-98-33}
\hbox{YCTP-P22-98}
\hbox{hep-th/9808060}
}}
{\vbox{\centerline{
Non-abelian Tensor-multiplet Anomalies}
}}

\centerline{Jeffrey A. Harvey}
\medskip
\centerline{Enrico Fermi
Institute and Department of Physics}
\centerline{University of
Chicago, 5640 Ellis Avenue, Chicago, IL 60637}
\bigskip
\centerline{Ruben Minasian and Gregory Moore}
\medskip
\centerline{Department of Physics, Yale University,}
\centerline{New Haven, CT 06520}

\bigskip
\centerline{\bf Abstract}

We use the anomaly cancellation of the M-theory
fivebrane to derive the R-symmetry anomalies of the
$A_{N}$ $(0,2)$ tensor-multiplet theories. This result
leads to a simple derivation of black hole entropy
in $d=4, \CN=2$ compactifications of $M$-theory.
We also show how the formalism of normal bundle
anomaly cancellation clarifies the Kaluza-Klein
origin of Chern-Simons terms in gauged supergravity
theories. The results imply the existence of interesting
$1/N$ corrections in the AdS/CFT correspondence.

\Date{August 7, 1998}

%
\newsec{Introduction}
Anomalies are related to topology.  Hence anomalous
couplings are robust and serve as effective probes
of the {\it terra incognita} of theoretical physics. In this
note we use anomalous couplings to learn about the
six-dimensional superconformal $(0,2)$ models with
nonabelian gauge symmetry. Our main result is equation
$(2.5)$. As a corollary, we easily recover the formula for
black hole entropy in M-theory found in  \msw.
Our main technique is similar to that used in a recent
discussion of the Chern-Simons term of
$D=11$ supergravity in the presence of 5-branes \fhmm.
Our considerations  turn out to be useful in explaining the
Kaluza-Klein origin
of Chern-Simons terms in gauged supergravity.
This is the subject of section
three. We also comment on a mismatch between anomalies in
the AdS/CFT correspondence and its implications.

\newsec{Anomalies of the $(0,2)$ theory}
\subsec{The anomaly polynomial}

The fivebrane of M-theory has chiral world-volume fields which
lead to potential anomalies in diffeomorphisms of the five-brane
world-volume $W_6$ as well as in diffeomorphisms which act as
$SO(5)$ gauge transformations of the connection on the normal bundle $N$.
The anomaly is determined by an eight-form $I_8$. There are two obvious
contributions to $I_8$. The  chiral world-volume fields of the fivebrane
lead to a contribution $I_8^{\rm zm}(Q_5)$. The second contribution arises
from anomaly inflow as a result of the coupling \refs{\vwrsq,\dlm}
\eqn\onea{\int_{M_{11}} C_3 \wedge I_8^{\rm inf}}
and leads to a contribution $I_8^{\rm inf}(Q_5)$.
For a charge $Q_5=1$ fivebrane these two contributions do
not cancel but rather \wittfive\
\eqn\oneb{I_8^{\rm zm}(1) + I_8^{\rm inf}(1) = p_2(N)/24.}

In \fhmm\ it was pointed out that there is a further contribution
to the anomaly which arises from a careful treatment of the Chern-Simons
term of $D=11$ supergravity in the presence of fivebranes. With
$\rho$ the integral of a bump form $d \rho$ and $e_3^{(0)}$
\foot{Throughout this paper we use the notation of the descent formalism,
$\omega_n=d\omega_{n-1}^{(0)}$, $\delta \omega_{n-1}^{(0)} =
d \omega_{n-2}^{(1)}$ for a closed gauge invariant n-form $\omega_n$}
related
to the global angular form $e_4/2$ by $e_4 = d e_3^{(0)}$ (see
Appendix A for details)  the
Chern-Simons term is
\eqn\zzi{S'_{CS} = \lim_{\epsilon \rightarrow 0} - {2\pi \over 6} \int_{M_{11}
-
D_{\epsilon}(W_6)} (\Csl_3 - \sigma_3) \wedge d(\Csl_3 - \sigma_3)
\wedge d(\Csl_3 - \sigma_3 ) }
where $\sigma_3 \equiv \rho e_3^{(0)}/2$. Here we have also removed a
tubular neighborhood of radius $\epsilon$ surrounding the fivebrane.
The variation of $S'_{CS}$ may be computed using a result of
Bott and Catteneo \bottcatt\ and leads to a contribution
\eqn\onec{I_8^{\rm CS} = - p_2(N)/24 }
which cancels the remaining anomaly.

The cancellation of all anomalies for a charge one fivebrane gives
us confidence that the anomalies must also cancel for arbitrary
charge $Q_5$ fivebranes. Unfortunately this is difficult to verify
explicitly because the world-volume theory for charge $Q_5$ is
not sufficiently well understood.
This theory is a non-Abelian tensor theory
with $(0,2)$ supersymmetry.

Instead,  we will
assume that the anomalies do in fact cancel for all $Q_5$
and use this to predict what
the anomalies of the non-Abelian $(0,2)$ theory must be. This is
easily done since the bulk supergravity
contributions from \onea\ and \zzi\ can be computed
reliably for $Q_5>1$. Note that
the contribution to the anomaly from $I_8^{\rm inf}$
is linear in $C_3$ hence linear in $Q_5$ while the contribution from
$I_8^{\rm CS}$ is cubic in $C_3$ and hence cubic in $Q_5$. Thus anomaly
cancellation requires that
\eqn\oned{I_8^{\rm zm}(Q_5) = Q_5 I_8^{\rm zm}(1) + (Q_5^3 -Q_5) p_2(N)/24, }
where \wittfive:
\eqn\defiate{
I_8^{\rm zm}(1) = {1 \over 48} \Biggl[ p_2(N) - p_2(TW) + {1 \over 4}
\left( p_1(TW)-p_1(N) \right)^2  \Biggr].
}
Here $TW$ denotes the
tangent bundle to the fivebrane worldvolume $W$, and $N$ is the normal bundle
to $W$ in the bulk $M_{11}$.

The extension of \oned\ to the $D$ and $E$ series of
$(0,2)$ theories and to the nonabelian $(0,1)$ theories
remains open.

\subsec{Application 1: Correlators in the $(0,2)$ theory}

Equation \oned\ contains some nontrivial information
about the current correlators of the $(0,2)$ nonabelian
superconformal theory of
nonabelian tensor-multiplets. This theory has an
$OSp(6,2\vert 4)$ superconformal current multiplet $\CJ$
whose structure is given  in part by
\hst:
$\CJ_{\rm bosonic} = (J^{(IJ)}, J_\mu^{[IJ]}, J_{\mu\nu\lambda}^{I},
J_{\mu\nu}) $, where
$I=1,\dots, 5$ is an $so(5) \cong usp(4) $
$\CR$-symmetry index in the fundamental
of $so(5)$, the scalar term $J^{(IJ)}$ is in
the ${\bf 14}$, $J_\mu^{[IJ]}$ are the $\CR$-symmetry
currents in the adjoint ${\bf 10}$, the anti-selfdual 3-form
currents $J_{\mu\nu\lambda}^{I}$ are in the ${\bf 5} $
and the energy-momentum tensor $J_{\mu\nu} $ is
a singlet. These currents can be coupled to
a contragredient multiplet of background
fields $\Phi_{\rm bosonic}= (\pi^{(IJ)}, A_\mu^{[IJ]},
S_{\mu\nu\lambda}^I, h_{\mu\nu}) $ to form the
generator of current correlators:
\eqn\generator{
\exp\biggl[ -\Gamma[\Phi] \biggr]  \equiv
\biggl\langle \exp\bigl[ \int_{W_6}  \Phi \cdot \CJ \bigr] \biggr\rangle
}
The result \oned\ for the anomaly polynomial
implies that if $\epsilon^{[IJ]}$ is an infinitesimal
$so(5)$ transformation then the normalized
correlator is:
\eqn\correlators{\eqalign{
\biggl \langle \biggl \langle\bigl( & \epsilon^{[IJ]} D_\mu J^{[IJ]\mu} \bigr)
 \exp\bigl[ \int_{W_6}  \Phi \cdot \CJ \bigr] \biggr\rangle
\biggr\rangle \cr
& = {Q_5^3 - Q_5 \over  24} \int_{W_6} (p_2)^{(1)}_6(\epsilon,A)
+ Q_5 \int (I_8^{\rm zm}(1))^{(1)}_6(\epsilon,A,h)
} }
Since
\eqn\twod{ p^{(0)}_2(A)  =  {1 \over 4} (  \half \omega_3^{(0)} \omega_4 -
\omega_7^{(0)} ) }
where $\omega_{2n}= ({i \over 2\pi})^n tr F^n$ and $\omega^{(0)}_{2n-1} =
d^{-1} \omega_{2n}$, we find results for
$4,5,6$ and $7$-point functions of currents.

The result \correlators\ implicitly contains a good deal of
information about correlators in the
$(0,2)$ theory.
In the $(0,2)$ theory the $\CR$-symmetry anomaly   is in the
same supermultiplet as the conformal anomaly as can be seen
by dimensionally reducing the theory to four dimensions.
Thus in principle \correlators\
provides an exact prediction of the conformal anomaly of the $(0,2)$ theory
after carrying out sufficiently
many supersymmetry transformations. The details
of this calculation are worth doing, but we have not
worked them out. The argument might be similar to
the discussion in appendix A of \afgj. Nevertheless,
without working out the details we may make some
qualitative observations.  The
$Q_5^3$ dependence of \oned\ is in accord with the expectations from
black hole calculations \klebs\ and with a large $Q_5$ calculation of the
conformal anomaly using the AdS/CFT correspondence \hennsken. The
novelty here is that we also predict an exact correction to the leading
$Q_5^3$ dependence which is down by $1/Q_5^2$.

Equation \oned\ provides some
 interesting clues to the structure of
the still unknown microscopic description
of the $(0,2)$ theory. For example,
it is quite likely that \oned\ and  \correlators\ can be used to
derive an infinite number of correlation functions when
the $(0,2)$ theory is compactified on 6-folds of
$SO(4) \times SO(2)$ holonomy. The strategy follows
the ideas of \clash. One would  start with the group cocycle class associated
with
\oned, then twist the theory to produce a scalar
supercharge $\CQ$ so that some of the currents
are $\CQ$-exact. The absence of
gauge fields contragredient to the $\CQ$-exact
currents in the effective action
then fixes the actual group cocycle representing
the class \oned. This  determination of the
``trivial cocycle'' fixes the kinetic terms in a
 generalized gauged WZW-type  action. The
resulting WZW-type action   then
serves as a generating function
for an infinite number of current correlators.

\subsec{Application 2: Black hole entropy}

A   precise check of \oned\ and its connection to the conformal
anomaly can be made by reducing the world-volume theory to a
$1+1$ dimensional conformal field theory. To do this we consider
wrapping a charge $Q_5$ M-theory fivebrane on a supersymmetric
four-cycle $P$ in a Calabi-Yau threefold $X$.

Let $\{ \theta_A \}$ be an integral basis for $H^2(X,Z)$ and denote
the dual basis of  $H_4(X,Z)$ by $\{ \sigma_A \}$. The three-form
potential of $D=11$ supergravity reduces to a set of $U(1)$ vector
fields through the Ansatz $C_3 = \sum_A C_1^A \wedge \theta_A$. A
single fivebrane wrapped on a smooth four-cycle in the
homology class $P_0 = P_0^A \sigma_A$ gives rise
to a string in $4+1$ dimensions which carries charge $P_0^A$ under
the $U(1)$ gauge field $C_1^A$. In what follows we write
$C_1$ for the linear combination of $U(1)$ gauge fields determined
by the element of $H^2$ dual to $P_0$.  The string has an $SO(3)$ normal
bundle and the zero mode and inflow contributions
to the anomaly fail to cancel by a term involving $p_1(N)$. As for
the fivebrane this anomaly was shown in \fhmm\ to be cancelled
by a modification of the Chern-Simons term of $D=5$ supergravity.

We now consider how the story changes when we scale
the homology class
$P_0 \rightarrow Q_5 P_0$.
Demanding that the anomaly be cancelled for charge $Q_5$ wrapped
fivebranes  predicts that the zero mode anomaly is given by
\eqn\onee{
I_4^{\rm zm}(Q_5)  = Q_5 I_4^{\rm zm}(1) + (Q_5^3-Q_5) p_1(N) D_0/4}
where \fhmm\
\eqn\oneex{
 I_4^{\rm zm}(1)  = {c_2 \cdot P_0 \over  48}    \bigl(
p_1(TW) + p_1(N) \bigr) +  {D_0 \over  4}  p_1(N) }
with
\eqn\inters{D_0\equiv {1 \over 6} \int_X \hat P_0^3 = D_{ABC} P_0^A P_0^B
P_0^C}
where $\hat P_0$ is the dual to $P_0$.

The string obtained from the wrapped fivebrane is described at
low-energies by a conformal field theory with $(0,4)$ supersymmetry.
The right-moving superconformal algebra has an $SO(3)$ Kac-Moody
algebra whose level $k$ is related to the $SO(3)$ normal bundle
anomaly above. Specifically, the level $k$ is equal to the
coefficient of $p_1(N)/4$ in \onee\ which gives
\eqn\onef{k = Q_5^3 D_0 + Q_5 c_2 \cdot P_0 /12}
The right-moving Virasoro central charge is given
in terms of the level of the $SO(3)$ Kac-Moody algebra
by $c_R = 6k$ so we see that
the conformal anomaly has terms cubic and linear in $Q_5$.

Let us now compare our result for
$c_R$ with the result in \msw.
We assume that $P_0$ is the divisor class of a line
bundle $\CL_0$ defining an embedding of $X$ into
projective space (that is, $\CL_0$ is ``very ample'').
In particular, the linear system $\vert\CL_0\vert$
has no basepoints (i.e. points where all sections
of $\CL_0$ vanish). Then, for $Q_5>0$ the linear system
$\vert Q_5 \CL_0\vert $ also has no basepoints and by Bertini's
theorem the homology class $Q_5 P_0$ has a representative
by a smooth 4-cycle.
\foot{See, \GrHa, p. 137, et. seq. for
a discussion of the relevant math.}  If the 5-brane is wrapped on
such a smooth  4-cycle the analysis of the zeromodes used in
\msw\ is justified, and leads to exactly the same
result, $c_R = 6 k$ with $k$ given by
\onef. This confirmation of a known result gives
us added confidence in \oned.

We would like to conclude this section with
an observation on black hole
entropy and anomaly inflow. As discussed in \msw, the microscopic
configuration of a fivebrane wrapping the five-cycle $P \times S^1$
and with momentum along the $S^1$ can be described in a certain
regime as an extremal black hole solution of $N=2$, $d=4$
supergravity. The entropy can be computed macroscopically in terms
of the area of the event horizon of the black hole and including
a one-loop topological correction. On the other hand the entropy
can also be computed by counting microscopic states in the
$(0,4)$ SCFT of the effective string.
This leads to an entropy
$S=2\pi \sqrt{{c_L q_0/6}} $, where $q_0$ is the total momentum
of the left-moving excitations and
$c_L$ is the left-moving Virasoro central charge.

The previous analysis
determined $c_R$ by cancellation of the normal bundle anomaly, but
$c_L-c_R$ and hence $c_L$ can be determined from cancellation of the
tangent bundle anomaly.
In particular, the tangent bundle anomaly on the worldsheet is
given in terms of $c_R-c_L$ by
\eqn\twodiman{(c_R - c_L) \left( {p_1(TW) \over 24} \right) }
This has to be compared with the anomaly inflow from variation
of the term
\eqn\grinterfive{\Delta S^5 = \lim_{\epsilon \rightarrow 0}
{c_2 \cdot P_0 \over 48}
\int_{M_{5}-D_\epsilon (W)}
C_1 \wedge   p_1(TM).}
coming from  reduction of $C_3 \wedge
I_8^{\rm inf}$ on the manifold $X$.
We thus see that  the cancellation of the tangent bundle anomalies
requires $c_L - c_R = Q_5 c_2 \cdot P_0 /2$.
This again is in agreement with
\msw, and gives
\eqn\centl{c_L = 6 Q_5^3 D_0 + Q_5 c_2 \cdot P_0 }
as expected. Thus both the left and right conformal anomalies of this
$(0,4)$ theory and hence the black hole entropy are completely determined
by anomaly cancellation. This explains in part
the results of \msw\ which reproduced the black hole entropy  precisely
without the need for string theory or a complete microscopic description
of M theory.

\newsec{Chern-Simons terms in Gauged Supergravity}
Gauged supergravity theories in odd dimensions typically contain
Chern-Simons couplings of the gauge fields. These have been found
in the literature using the Noether method to determine the supersymmetric
completion of the Einstein action. Recently these Chern-Simons terms
have been of interest in connection with the AdS/CFT correspondence
\refs{\largen, \gkp,\wittads}. Indeed, following
\wittads\ we identify the background fields
$\Phi$ of \generator\  with the
``boundary values'' of the
$AdS_7$ maximally extended supergravity multiplet
and $\Gamma[\Phi]$ as the on-shell supergravity
action (suitably regularized). In particular, it follows
 \wittads\ that the anomaly on the boundary of AdS space
associated with the variation of the Chern-Simons terms should match
the anomaly computed in the boundary CFT.

Since gauge supergravities arise from compactification of higher dimensional
theories on compact spaces with isometries (typically spheres in the
simplest examples) it must be possible to understand the Chern-Simons
terms from   Kaluza-Klein reduction
\refs{\acf, \kkrefs}. Unfortunately, such
 reductions are notoriously subtle,
and to our knowledge have not been carried out in the literature to
the non-linear order necessary to see the Chern-Simons terms.
We will show here that the formalism necessary for smoothing out brane
sources has a direct application to this problem. We then comment on
the matching of anomalies.

\subsec{Chern-Simons terms in $AdS_7$}

Seven-dimensional supergravity in $AdS_7$ with gauge group $SO(5)$
arises by Kaluza-Klein compactification of M-theory on $S^4$ \freund.
The vacuum configuration is given by the standard metrics on the
maximally symmetric space $AdS_7 \times S^4$ and
\eqn\twoa{\Gsl_4  = Q_5 \epsilon_4}
with $\epsilon_4$ the volume form on $S^4$ and $\Gsl_4 = G_4/2 \pi$.
In order to carry out
a Kaluza-Klein reduction it is necessary
to expand the metric and
four-form field strength to include fluctuations.
This was done at the linearized level in \pvnt, but
the extension to the nonlinear theory is not obvious.
The required ansatz is highly constrained by
the requirements that $G_4$ be gauge invariant under $SO(5)$
gauge transformations, that the Bianchi identity $d G_4=0$ be
satisfied, and by the requirement that $\int_{S^4} \Gsl_4/2  = Q_5$. The
formalism used in \fhmm\ is well-suited to
finding such an ansatz.

In the presence of fluctuations of the $SO(5)$ Kaluza-Klein gauge
fields the ansatz \twoa\ can be made gauge covariant by replacing
ordinary derivatives with
covariant derivatives but it is  then  not
closed without the addition of extra terms. The modifications
needed to make it closed while maintaining gauge invariance
were derived in \fhmm\ and lead to the ansatz
\eqn\twob{\Gsl_4 = Q_5 e_4(A)/2 + {\rm fluctuations~~~ in ~~ C_3}}
where   $e_4(A)$ depends on both the metric
and the $SO(5)$ gauge field $A$. An explicit
expression is given in the appendix.
Keeping only terms involving the $SO(5)$ gauge fields the
Kaluza-Klein reduction of the Chern-Simons term then gives
\eqn\twoc{ \eqalign{-{2 \pi  \over 6} \int_{M_{11}} \Csl_3 \wedge \Gsl_4
\wedge \Gsl_4 &= -{2 \pi Q_5^3   \over 6} \int_{M_{11}} {e_3^{(0)} \over 2}
\wedge {e_4 \over 2} \wedge {e_4 \over 2}  \cr
&= - { 2 \pi Q_5^3 \over 24}
\int_{AdS_7} p^{(0)}_2(A) }}
where in the last step we have used the Bott-Catteneo formula.
Using \twod\ above we may compare with the
expressions in the supergravity literature \ppvn\ and
we find agreement.

In addition to the order $Q_5^3$ Chern-Simons term \twoc\ there must
also be additional Chern-Simons terms which are linear in $Q_5$
and which follow from Kaluza-Klein reduction of the $C_3 \wedge I_8^{\rm inf}$
term of $D=11$ SUGRA. Note that these terms would not have appeared
in earlier treatments because the $C_3 \wedge I_8^{\rm inf}$ term does not mix
under supersymmetry with the Einstein term but only with other
higher dimension terms in the M theory effective
action \refs{\greenco, \afmn}. If we restrict attention
to the leading order in a derivative expansion the
Kaluza-Klein reduction is easily carried out as follows.
The integration over $S^4$ requires a factor of the
volume form. We extract this term from $e_4(A)$
(the other terms contribute to higher derivative interactions).
Next we observe that, {\it to leading order in the
derivative expansion} the spin connection in the
standard Kaluza-Klein ansatz, restricted to a
cross section of the trivial bundle
$AdS_7 \times S^4 \rightarrow AdS_7$ is
simply a direct sum connection
\foot{See, e.g., equation $(A.21)$ in
\kkrefs.  }
\eqn\drctsum{
\omega_{AB} = \pmatrix{ \omega_{\alpha\beta} & 0 \cr
0 & (\nabla_a K^{[IJ]}_b ) A^{[IJ]} \cr} + \cdots
}
on $AdS_7$ for
$T(AdS_7) \oplus E$ where $E$ is a
vector bundle associated to the  $SO(4)$
tangent space group of $S^4$, and restricted to
$AdS_7$. In \drctsum\ the indices $A,B$ are eleven-dimensional tangent
space indices, $\alpha,\beta$ are
$AdS_7$ tangent space indices, and
$a,b$ are $S^4$ tangent space indices. Also,
$I,J$ are indices in the fundamental of $SO(5)$
and $K^{[IJ]}_b$ are components of Killing
vectors on $S^4$.
Equation \drctsum\ is an identity for  1-forms on
$AdS_7$.

Now, $I_8^{\rm inf} = (p_2 - p_1^2/4)/48$,
so we may use the formula for the total
Pontryagin class of a direct sum $p(E \oplus F) = p(E) p(F)$,
which is valid at the level of forms for a direct sum
connection, to obtain the Kaluza-Klein reduction on
$AdS_7$ to leading order in derivatives:
\eqn\coups{2 \pi Q_5 \int_{AdS_7}
\Biggl\{ (I_8^{\rm inf})^{(0)}_7(R) - {1 \over 48}
\Biggl[{p_1^2(A) \over 4} - p_2(A) -{ p_1(R) p_1(A) \over 2} \biggl]^{(0)}_7 +
... \Biggr\} }
Where now $p_i(R)$, $p_i(A)$ are representatives of the
Pontryagin classes of the seven-dimensional tangent bundle and
$SO(5)$ principal bundle respectively.
Supergravity in $AdS_7$ thus contains both the order $Q_5^3$
Chern-Simons term
\twoc\ and the order $Q_5$ terms \coups.

Similar techniques should allow a direct Kaluza-Klein derivation of
Chern-Simons terms in other cases of interest such as IIB
SUGRA on $AdS_5 \times S^5$. In this case there are
 several additional subtleties. These include   the lack
of a covariant action for the self-dual five-form of IIB theory and
the fact that in the Kaluza-Klein reduction the massless
$SO(6)$ gauge multiplet is in fact a linear combination
of a Kaluza-Klein mode of the ten-dimensional metric   and a
Kaluza-Klein mode of the
IIB 4-form potential
\krvn.

\subsec{Implications for the AdS/CFT Correspondence}

The AdS/CFT correspondence requires matching of the anomalies computed
in the CFT and from variation of the Chern-Simons terms in AdS
supergravity. This matching has been verified in detail at
large $N$ \cnss.

Since the anomaly is exact, we can also ask whether the anomaly
matches at subleading order in $N$. Consider for example the
correspondence between $N=4$ SYM and supergravity on $AdS_5 \times S^5$.
Classical supergravity predicts the Chern-Simons
term has coefficient $N^2$. On the other
hand, the exact coefficient can be computed from
anomaly inflow arguments. Since the gauge multiplet
on the D3 branes is $SU(N)$, and {\it not}
$U(N)$  \refs{\wittads, \ahaw}, the exact  coefficient of the Chern-Simons
term must in fact be proportional to $N^2-1$.
If the correspondence is correct then there must be a
correction to the Chern-Simons term (and by supersymmetry a correction
to the Einstein term as well) of order $1/N^2 \sim g_s^2 {\alpha'}^4$.
Such a one-loop higher derivative correction does not seem
to be ruled out by any known renormalization theorems.
Confirming this correction would be a non-trivial check of
the correspondence at finite $N$ and therefore of
the correspondence in one-loop string theory and not just
in the classical supergravity limit. Similar comments apply to
the $AdS_7 \times S^4$ case where again there must be an order
$1/N^2$ correction to the leading $N^3$ behavior (in addition to
the order $N=Q_5$ term of \coups).

\bigskip
\centerline{\bf Acknowledgments}\nobreak
\bigskip

We would  like to
thank O. Aharony, T. Banks, S. Ferrara, M. Green,
K. Pilch,   S. Shatashvili, P. van  Nieuwenhuizen,    and
particularly E. Witten for
discussions. GM would like to acknowledge the hospitality of
the Aspen Center for Phyiscs, and GM and JH
acknowledge the hospitality of the Amsterdam Summer
Workshop on String Theory and Black Holes.
The  work  of JH is supported by NSF Grant No.~PHY 9600697, RM and GM are
supported by
DOE grant DE-FG02-92ER40704.

\appendix{A}{Non-singular branes}

Consider a $p$-brane  with worldvolume $W_d$ (the longitudinal coordinates are
$x^{\mu}$, $\mu = 0,1, \ldots d=p+1$) located at $y^{a}=0$, $a=1,2, \ldots D-d$
in the total space $M_D$. The most naive expression for the Bianchi identity in
the presence of the brane (the magnetic source equation) is
\eqn\seven{d G_{D-d-1} = 2 \pi
\delta(y^1) \cdots \delta(y^{D-d}) dy^1 \wedge \cdots
\wedge d y^{D-d} .}
The quantity on the right hand side is a $(D-d)$-form with integral one
over the transverse space and delta function support on the brane.
In order to have a completely well defined  and non-singular
prescription in such cases we need to smooth out the delta function source.
Having done this we will see that
in the presence of a non-zero $SO(D-d)$ connection on the normal bundle
we will have to modify  the right hand side
 of \seven\ in order that it
transform covariantly under $SO(D-d)$ gauge
transformations.

This has been done in \fhmm\ for the case of the $M$-theory fivebrane. After
defining  a radial
direction away from the brane,  we cut out a disc of radius
$\epsilon$ around it. That is, we remove a tubular neighborhood of
the brane of radius $\epsilon$, $D_\epsilon(W_d)$. The map $\xi: D_\epsilon
\rightarrow W_d$ is a fibration with the fiber being an open $(D-d)$-ball which
may be considered as the unit ball
in the normal bundle.
The
restriction of $\xi$ to the set of points, $S_{\epsilon}(W_d)$, whose distance
to $W_d$ is fixed (and smaller than $\epsilon$) can be identified with a unit
sphere in the normal bundle   and thus has fiber $S^{D-d-1}$.
Note that $S_{\epsilon}(W_d)$ is the boundary of the tubular neighborhood
$D_\epsilon(W_d)$.

In order to smooth out the brane source we choose a smooth function of the
radial direction with transverse
compact support near the brane, $\rho(r)$,  with
$\rho(r) = -1$ for sufficiently small $r$ and
$\rho(r) =0$ for sufficiently large $r$.
The bump form $d \rho$ then has integral  one in the
radial direction.
The smoothed form of \seven\ should then read
\eqn\nine{ d G_{D-d-1} = 2 \pi \Phi_{D-d}, }
where $ \Phi_{D-d}$ represents the Thom class of the normal bundle. One can
write this representatitive
in terms of the global angular form and $\rho$. The expression depends on
whether the rank of the bundle is even or odd and is given by:
\eqn\thom{\eqalign{\Phi_{D-d} &= d \rho \wedge e_{2n}/2,
\,\,\,\,\,\,\,\,\,\,\,\,\,\,\,\, 2n=D-d-1   \cr
&= d (\rho \wedge e_{2n-1}/2), \,\,\,\,\,\, 2n-1=D-d-1.}}
 The global angular form $e_{D-d-1}$ is gauge invariant under $SO(D-d)$
transformations
of the normal bundle. $\Phi_{D-d} $ should reduce to the naive expression
on the r.h.s of \seven\ for a flat infinite fivebrane when $d \rho$
approaches a delta function. Physically what we are doing
is smoothing out the magnetic charge of the brane to a
sphere of magnetic charge linking the horizon.

We now give explicit formulae for the global angular form
on a real   vector bundle  $E \rightarrow M$   with metric and connection.  Let
$E_0$ be the complement of the
zero-section.

If  ${\rm rank}(E) = 2n+1$ is odd then
the sphere bundle has fibers $S^{2n}$.
The  global angular form
$e_{2n}$ on $E_0$ restricting to the volume form
on the fibers of $S(E)$ satisfies
\eqn\eulclssx{
d e_{2n} =0
}
so the Euler class vanishes:
\foot{This equation holds rationally. In fact, one only
needs to be able to invert $2$.}
$\chi(E) =0$. Moreover,  $e_{2n}/2$ has integral one over
the fibers of
$S_{\epsilon}$. The global angular form is given by
\eqn\explcten{
\eqalign{
e_{2n}=  {1 \over  2 (4\pi)^n n! }
\sum_{j=0}^{n} (-1)^j {n! \over j!(n-j)!}
\epsilon_{2n+1} (F)^j (D\hat y)^{2n-2j}  \hat y, \cr}
}
where  $\hat y^{\hat a} \equiv  y^{\hat a}/r$ and are defined only outside of
$0\in
\IR^{D-d}$ (${\hat a}= 1, \ldots, D-d$). There is a globally-defined connection
$\Theta$ on the total space
of the  $SO(D-d)$
bundle in terms of which we have
\eqn\horizontal{
\eqalign{
(D \hat y)^{\hat a} & \equiv d\hat y^{\hat a} - \Theta^{{\hat a}{\hat b}} \hat
y^{\hat b}\cr
F^{{\hat a}{\hat b}} & = d \Theta^{{\hat a}{\hat b}} - \Theta^{{\hat a}{\hat
c}} \wedge \Theta^{{\hat c}{\hat b}}, \cr}
}
and
\eqn\ixdfh{\epsilon_{2n+1} (F)^j (D\hat y)^{2n-2j}  \hat y
\equiv \epsilon_{{\hat a}_1 \ldots {\hat a}_{2n+1} } F^{{\hat a}_1 {\hat a}_2}
\ldots F^{{\hat a}_{2j-1} {\hat a}_{2j}}
(D\hat y)^{{\hat a}_{2j+1}}   \ldots(D \hat y)^{{\hat a}_{2n}} \hat y^{{\hat
a}_{2n+1}}.}
\noindent The cohomology class $e_{2n}$ satisfies \bottcatt\
\eqn\etwon{[e^2_{2n}]=\pi^{\ast}(p_n(E)),}
Moreover, at the level of differential forms we have
 \eqn\bottca{\pi_{\ast}(e_{2n}^3) = \pi_{\ast}\left( e_{2n} \pi^{\ast}p_n
\right) = 2p_n
}
for the expression above.

If  ${\rm rank}(E) = 2n$ is even   the sphere bundle has
fibers $S^{2n-1}$.  There is a global angular form
$e_{2n-1}$ on $E_0$ restricting to the volume form
on the fibers of $S(E)$ such that
\eqn\eulclss{
de_{2n-1} = - \pi^*(\chi(E))
}
for  $\chi(E) \in H^{2n}(M;\IZ)$.  The global angular form is given by
\eqn\explct{
\eqalign{
e_{2n-1}=  -{1 \over   (2\pi)^n }
\sum_{j=0}^{n-1} {2^{-j} \over j!(2n-2j -1)!!}
\epsilon_{2n} (F)^j (D\hat y)^{2n-2j-1}  \hat y. \cr}
}

\listrefs
\bye